\documentstyle[11pt,aaspp4]{article}
\begin{document}
\title{Dust lanes causing structure in the extended
narrow line region of early-type Seyfert galaxies}

\author{
A.~C.\ Quillen\altaffilmark{1}$^,$\altaffilmark{2},
A.\ Alonso-Herrero$^1$,
M.~J.\ Rieke$^1$,
Colleen McDonald$^1$,
Heino Falcke$^{1,}$\altaffilmark{3}, \&
G.~H.\ Rieke$^1$
}
\altaffiltext{1}{The University of Arizona, Steward Observatory, Tucson, AZ 85721}
\altaffiltext{2}{E-mail: aquillen@as.arizona.edu}
\altaffiltext{3}{Max-Planck-Institut f\"ur Radioastronomie, Auf dem H\"ugel 69, D-53121 Bonn, Germany}

\begin{abstract}
We construct near-infrared to visible broad band NICMOS/WFPC color maps
for 4 early-type Seyfert galaxies with S-shaped or one-sided
ionization cones.   We find that dust lanes are
near or connected to many of the features seen in the [OIII]
and H$\alpha$+[NII] line emission maps.  This suggests that much of the
structure of line emission in these ionization cones is determined by
the distribution of ambient dense galactic gas.  
Spiral arms, dust lanes caused by bars, or gaseous warps
provide dense gas which when illuminated by a conical beam 
of ultraviolet photons 
can result in the complicated line emission morphologies observed.
\end{abstract}


\section{Introduction}

In Seyfert galaxies when radiation is obscured by an inner optically thick
`torus' (\cite{antonucci}) ultraviolet (UV) radiation can
escape along a conical shaped beam 
causing an ``ionization cone'' to be observed in emission line maps
(e.g., \cite{evans}, \cite{pogge89}).
The observed morphology and luminosity of the ionization cone can 
be influenced by the density distribution of the ambient media 
(e.g., as simulated by \cite{mulchaey96}).
For example, in NGC~4151 extended emission is produced  
by the intersection of the ionization
cone with the disk of the galaxy (\cite{pedlar}; \cite{boksenberg};
\cite{wilson+t}).
In samples of Seyfert galaxies 
the orientation of [OIII] and H$\alpha$+[NII] line emission 
is generally near that of the galaxy major axis
(\cite{nagar}, \cite{mulchaey95}).
This suggests that the availability and distribution of dense galactic gas
plays an important role in determining the width  
and orientation of the ionization cone.

Extended emission-line and radio morphologies are often co-spatial or
aligned in Seyfert galaxies (e.g., \cite{unger}; \cite{pogge}).
This supports the idea that ionizing photons
preferentially escape along the radio axis.
The connection between the radio ejecta of Seyfert nuclei
and their extended emission line regions 
is evident from the similar spatial extents
observed in high angular resolution radio interferometric
observations compared to Hubble Space Telescope (HST) images
(e.g., recently \cite{simpson}; \cite{falcke3}; \cite{mulchaey94};
\cite{ferruit98_3}).  However, the morphologies observed 
in lines from ionized gas [OIII]$5007\AA$ and H$\alpha$+[NII]$6548, 6583\AA$
can be complex, showing S-shaped features, partial loops, and curves 
suggestive of bow shocks.  Except for
observations showing that high excitation gas tends to form a
cone shaped morphology, no pattern has emerged 
connecting the complicated morphologies observed in the
ionized gas distribution among different Seyfert galaxies. 
In particular it not clear what role the distribution
of ambient galactic gas plays compared to energetic hydrodynamical
processes caused by the jet.

In this paper we take the opportunity provided by 
high angular resolution near-infrared imaging with NICMOS on HST 
to make color maps with
previously observed visible wavelength HST/WFPC2 images.  
We find that the resulting color maps show extinction from dust features
with higher signal to noise 
than possible with color maps made previously from the existing 
WFPC2 images alone.
This could be because the large wavelength separation 
between the visible and near-infrared images 
makes it easier to identify features with low column depths of dust.
We present here a comparison between morphology observed
in these color maps (tracing dust features) 
and that displayed in [OIII] and H$\alpha$+[NII] line emission from ionized gas.
Using these color maps we can examine how the distribution
of ambient dense gas in the galaxy affects the morphology
of the line emission.

We chose Seyfert galaxies with existing high quality images
observed with WFPC or WFPC2 and NICMOS 
%(cameras on board HST) 
and published recent literature discussing their
ionization cone morphology.  The galaxies are 
also of early-type, so as to minimize 
confusion caused by star formation and by extreme extinction from dust.
This limited our
sample to 4 galaxies: Markarian~573, NGC~3516, NGC~2110 and
NGC~5643.  All are Seyfert 2 galaxies except for NGC~3516
which is a Seyfert 1.  The Hubble types range from SB0 to SAB.

A number of dynamical models or scenarios have been proposed
to explain the individual galaxy morphologies.
In NGC~3516 a bent bipolar mass outflow model was suggested
(\cite{goad}; \cite{mulchaey92}) to account for the S-shape of
the ionized gas.  Alternatively
a precessing jet model might also be appropriate (\cite{veilleux}).
In Markarian~573 inner knots nearest the radio jet may represent 
deflection of the jet by ambient or entrained clouds, whereas the inner
arcs at $1''.8$ may represent bow shocks driven
by the jets (\cite{ferruit99}).    Despite the similarities in morphology
between the inner and outer arcs at $3''.6$ this mechanism
is unlikely to apply to the outer arcs.
In NGC~2110 a pair of curved features of ionized gas is observed which
are offset from the radio jet itself (\cite{mulchaey94}). This
might be consistent with a model where 
gas is swept up or ejected by the jet,  but the relative positions
of the radio emission and the emission line features remains 
difficult to explain (\cite{mulchaey94}).
Despite the morphological similarity between the inner and
outer emission features in NGC~2110
the outer S-shaped region of ionized gas at $4''$ from the nucleus
is suspected to be caused by a different
mechanism, that of ambient interstellar gas photoionized by
the central source (\cite{mulchaey94}).
Many of these studies and interpretations were hampered by the
lack of information about the distribution of ambient galactic material.
In this paper we search for a simple explanation for the variety
of emission line morphologies observed in these galaxies.

%While the radio and emission line morphologies are generally nearly
%co-spatial, high resolution observations have shown that
%there are significant differences in alignment.  For example,
%in Markarian~573 and NGC~5643 the radio emission
%is observed to be to one side (but still within) the region
%of [OIII] or H$\alpha$+[NII] emission (\cite{simpson}, \cite{ferruit99}).
%In NGC~2110 the radio emission is almost anticorrelated with 
%regions of bright [OIII] and H$\alpha$+[NII] emission (\cite{mulchaey94}).

%Seyfert types:  
%NGC~3516      Sy 1  SB0  2649         171pc/''  D=35.3Mpc
%NGC~5643      Sy 2  SAB  1199         77pc/''   D=16.0Mpc
%Markarian~573 Sy 2  SAB0 5174km/s     335pc/''  D=69.0Mpc
%NGC~2110      Sy 2  SAB0 2284         148pc'''  D=30.4Mpc

\section{Comparison of near-infrared/visible color maps with emission
line maps}

WFPC, WFPC2 (visible) and NICMOS (near-infrared) 
images were taken from the HST archive.
For more information on the visible band images see
the original papers discussing the HST observations 
(on NGC~3516; \cite{ferruit98},
on NGC~5643, \cite{simpson}, on NGC~2110; \cite{mulchaey94}, 
on Markarian~573; \cite{ferruit99}, \cite{falcke}).
NICMOS Camera 2 images in the filter F160W (centered at $1.60\mu$m)
were primarily from the snap shot program 7330 (\cite{regan_mul}).
For NGC~2110 
%(black listed and then subsequently not observed by the NICMOS/GTO program) 
we used the narrow band image in the filter F200N (centered
at $2.00\mu$m) from GO program 7869.  
The NICMOS images were reduced with nicred (\cite{mcl})
with on-orbit flats and darks.
In Figures 1-4 we show ionized gas traced in either [OIII]
or in H$\alpha$+[NII] for the four galaxies. 
We also show color maps constructed from NICMOS and WFPC2 images.
The color maps trace the morphology of molecular gaseous structures
predominantly on the near side of the galaxy (in front
of most of the stars).  They are the only way currently feasible
to compare the location of the molecular gas with the emission line 
structures in the ionization cones.

\subsection{NGC~5643}
The isophotes in the outer parts of the NICMOS camera 2 F160W, $1.6\mu$m  
image are roughly aligned with a large scale bar
(with major axis $PA\sim 90^\circ$, 
and extending to a radius of $r_b\sim 30''$, \cite{regan_3}).  
No inner bar is detected in this image.
The color map shows a pair of extinction features consistent with leading
dust lanes along but offset from the major axis of the bar.
Knots observed in [OIII] are probably not directly
associated with any dust features observed in the galaxy.
However, the southern side of the ionization cone displays a component 
of diffuse emission which appears to be bounded by the 
dustlane seen on the south east side of the galaxy.  
The dust lane appears to be very slightly offset to the south from the 
diffuse component of the line emission.  
This could be explained by a projection effect if the UV 
radiation beam from the central source (or from shocks caused by the jet, 
e.g., \cite{dopita}) illuminates material
somewhat above but not in the plane of the galaxy.
This offset could also be explained by a model where
dense material originally from the dust lane is entrained 
by moving material associated with the jet ($PA \sim 90^\circ$, \cite{morris}).

\subsection{NGC~2110}
For NGC~2110 the deepest visible broad band image available
was the F606W broad band filter which is somewhat contaminated by 
line emission.  However the extent of dust lanes can be seen
outside the ionization cone 
and includes large areas which are not contaminated.
The overall pattern is that of spiral dust lanes which could
be part of spiral arms.
In particular dust features are observed at about $4''$ and $8''$ north
of the nucleus, roughly corresponding to the two arcs seen in 
the H$\alpha$ and [OIII] emission maps (\cite{mulchaey94}). 
A dust lane is also observed
to the south of the nucleus corresponding to a broad curve of line emission
about $2''$ south of the nucleus.  The dust features appear to be 
offset from the line emission, being slightly more distant
from the nucleus (see the above discussion on NGC~5643).  

\subsection{NGC~3516}
On large scales NGC~3516 is barred with major axis $PA\sim 170^\circ$  
and extending to a radius of $r_b\sim 13''$ (\cite{regan}).  
In the F160W image the isophotes are slightly elongated in a direction
roughly perpendicular to this outer bar (at $r \sim 3''$) and so
the galaxy may be doubly barred.  However the morphology of the 
dust features observed in extinction in the color map
and gas kinematics are
not consistent with what would be expected from gas in the plane
of the galaxy (see discussion in \cite{ferruit98}, \cite{veilleux}).
When gas exists above the plane of galaxy a warped configuration
is most likely (e.g., \cite{tubbs}).
To the south  of the nucleus a curved dust feature is observed in the color map.
The shape of this dust feature corresponds
quite well with the morphology of the [OIII] emission and suggests
that dust is associated with the ionized gas.
To the north west and south east of the nucleus extinction features
are observed which are not co-spatial with bright line emission.
We note that 
the F547M image we used to make the color map should be free of line emission.

\subsection{Markarian~573}
On large scales Markarian~573 is barred with major axis $PA\sim 0^\circ$  
and extending to a radius of $r_b\sim 10''$ (e.g., \cite{almudena}).  
In the F160W image there is a strong elongation almost exactly
perpendicular to the larger scale bar at $r \sim 2''$ which
corresponds to an inner bar (noted by \cite{pogge93}).
There is also a pair of dust lanes slightly offset from
the major axis of this inner bar (\cite{pogge93}) 
which would be consistent with leading dust lanes along this bar, 
however a warped dusty disk oriented perpendicular to the jet might
also be present.
% pa of major axis of galaxy on large scales seems to be 55 Adams 1977
% this galaxy has an outer ring

Unfortunately the F606W filter is strongly contaminated by line emission.  
However, in the F606W/F160W color map we can trace the extent of
dust features outside the ionized gas.  To the east of the nucleus
2 linear dust features 
are connected to the 2 south eastern arcs of line emission suggesting
that they are likely to continue within the region of 
line emission.   To the south there is also a dustlane
which connects to the outer arc.
The morphology of the line emission and dust lanes
is reminiscent of double spiral arms which are 
sometimes observed at the ends of a bar 
(e.g., in NGC~1365, \cite{lindblad}).   
The linear feature of high excitation emission
within an arcsec from the nucleus 
does not correspond to any dust features
and so is probably directly associated with the jet (as discussed
in \cite{ferruit99}, \cite{falcke}).

\subsection{Interpretation}

Spiral arms and bars can cause gas mass surface density contrast
ratios (between arm and inter-arm of dustlane and inter-dustlane)
of greater than a factor of few in the plane of the galaxy
(e.g., \cite{hau84}, \cite{ath92}).
Extinctions measured from our color maps are similar to that
estimated from visible color maps and 
range from $A_V \sim 0.5 - 1.5$ mag in the dust features 
(e.g., \cite{simpson}, \cite{mulchaey94}, \cite{ferruit98})
corresponding to $N_H \sim 3-9 \times 10^{21}{\rm cm}^{-2}$
using a standard ratio of total hydrogen column depth to color excess.
Outside the dust features visible to infrared colors 
match those on the opposite
side of the galaxy nucleus allowing us to limit $A_V \lesssim 0.1$ mag
inter-arm (or inter-dustlane).
This implies that the minimum surface density contrast ratio
between arm/inter-arm (or dustlane and inter-dustlane)
is a factor of a few.  
The dust features that are evident in our visible/infrared 
color maps therefore represent a significant source of dense
galactic gas compared to that outside these features.

Densities estimated from emission line diagnostics are relatively
high ($\sim 30 ~ {\rm cm}^{-3}$ for NGC~3516, \cite{ulrich}; 
$100-850  ~ {\rm cm}^{-3}$  depending on the arc in Markarian~573, 
\cite{ferruit98}; $\sim 50 ~ {\rm cm}^{-3}$  
in a particular knot in NGC~5643, \cite{simpson}).
Using these densities to estimate the total number of ionized 
hydrogen from the H$\alpha$ line strength we estimate that
$N_{H,ion} \sim$ a few $\times 10^{21}$ cm$^{-2}$ in Markarian~573, 
NGC~2110 and NGC~3516 and a factor of 10 lower in NGC~5643.
We estimate that these column depths are a significant fraction
of those estimated above from the color excesses or extinctions
in the dust features.
The high densities and large column depths of ionized
hydrogen suggest that a nearby source of dense galactic
gas is required to account for their existence.
The dust lanes provide a nearby source of dense gas
which does not exist outside these regions.

The emission measure of a line is 
proportional to the density squared and so is a strong function of 
the gas density.  This implies that denser material should produce brighter
line emission.  If UV photons preferentially escape 
along a conical radiation beam, the densest material
illuminated by this beam would be easiest to see in a line emission map.
Alternatively in models where the interaction of the radio
ejecta and the ambient medium could also produce ionizing 
radiation (e.g., \cite{dopita}) we would also expect a bias 
towards detecting denser gas associated with dust features.  
When dense gas exists in the plane of the galaxy we preferentially
expect to see line emission near this plane, and associated
with denser media which would be traced by dust lanes.

%We now consider the possibility that
%the features we see in the color maps might be caused
%by an emission source important at 1.6$\mu$m (or $2.00\mu$m in the case
%of NGC 2110).  In NGC 2110 and Markarian 573 
%we observe linear red features in the color maps which
%are smoothly connected to particular linear line emission features. 
%Few other processes could result in red
%colors to be observed near ionized gas; (emission from
%dust would require $T> 2000K$, and emission lines in the near
%infrared bands such as [FeII], are typically low in equivalent width).
%It is very unlikely that such processes
%would be important outside the ionization cone.
%We can therefore regard the red features observed in the color
%maps outside the location of the ionized gas as most likely
%being caused by extinction from dust.
%Since it is unlikely that dust lanes are sharply truncated,
%they should continue into the region where the ionized gas is observed.

\section{Discussion }

For 4 early-type Seyfert galaxies we demonstrate that dust lanes are 
near features seen in line emission maps 
in the extended emission line region.
Although the correspondence between emission line and dust
features is not perfect, it is plausible that ``missing''
dust features that are seen in line emission lie on the
far side of the galaxy and are not seen as shadows in the color maps.
In NGC~5643 a component of diffuse line emission is bounded
by a dustlane on the south eastern side of the ionization cone.
In NGC~2110 3 spiral dust lanes have similar curvature
and location to 3 arcs seen in line emission maps.
In Markarian 573 on the east side 2 linear dust lanes
merge into the two south eastern line emission arcs.
To the south a linear dustlane merges into the south eastern outer arc.
In NGC 3516 to the south west of the nucleus patchy dust features exist
near the location of bright line emission.  The dust features continue
outside the region of ionized emission to the north west and
south east of the nucleus.

The proximity of dust features with line emission 
suggests that the morphology of the line emission
is affected by spiral arms or bars 
(except for NGC 3516 where the dust may be part of a gas warp). 
Dense gas in these dust features is more likely
to result in bright line emission when illuminated
by a UV source or when affected by a jet.
This suggests that line emission luminosity might be
dependent on jet or ionization cone orientation with respect
to the galaxy plane, or on the scale height of the galactic
gas.

When spiral arms or bars are present, moderate deviations from circular motion
caused by streaming are expected.  
This could explain some (but not all) of the peculiarities observed
in the velocity fields (e.g., in Markarian 573; \cite{ferruit99}).
However, we note that not all of the line emission features are 
associated with dust lanes.
This is illustrated with Markarian~573 and NGC~5643 where many of the features
observed in the [OIII] emission map do not have nearby
dust lanes.  We would expect that galaxies with a larger fraction of 
active star formation than the galaxies presented here
may have much more complicated ionized gas structure 
because of their multiple sources of UV radiation 
(e.g., as seen in Circinus, \cite{circinus}).
The galaxies studied here have
ionization cones which extend hundreds of pc from the galaxy nucleus, 
and well outside a gas disk exponential scale length.
Ionization cones on small scales within a gas exponential length 
would not be expected to be so sensitive to the distribution
of galactic gas in the plane of the galaxy.

%Another consequence of the association pointed out here
%involves interpretation of narrow line emission fluxes
%in the context of unification models.  If the ionization cone
%must be pointed near the galactic plane in order for
%[OIII] line emission to be bright, then there is a bias towards
%detecting narrow line emission in galaxies with opening axes or 
%jets oriented near the plane of the galaxy.   
%This would primarily affect
%interpretation of unification models if extinction on large
%scales, (e.g., 100 pc as compared to 1pc in the 
%molecular torus model) from the galaxy itself affects the 
%classification of Seyferts.  We might also expect
%orientation angles to be correlated with Seyfert luminosity
%as measured from the [OIII] line flux.

\acknowledgements
Support for this work was provided by NASA through grant number
GO-07869.01-96A 
from the Space Telescope Institute, which is operated by the Association
of Universities for Research in Astronomy, Incorporated, under NASA
contract NAS5-26555.
We also acknowledge support from NASA projects NAG5-3042 and NAG5-3359.
We acknowledge helpful discussions and correspondence with  
Chien Peng and Don Garnett.
%We thank the referee for many comments which have improved this paper.

\clearpage

%References

\clearpage

\begin{figure*}
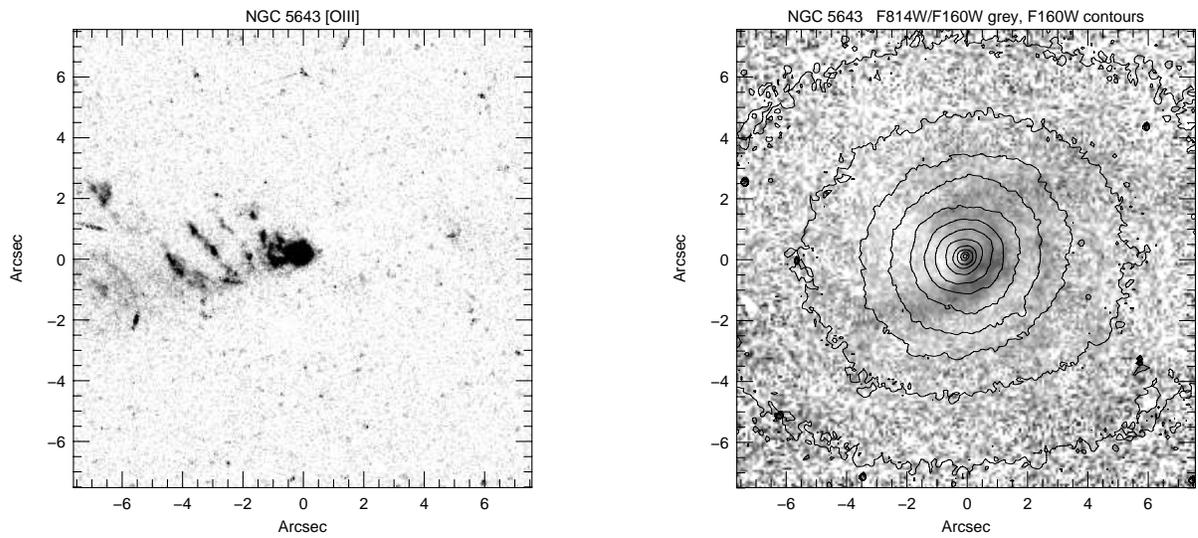

\caption[junk]{
a) The NGC 5643 [OIII] emission map.
b) The NGC 5643 color map (shown as the greyscale image).
Darker grey corresponds to redder colors or more extinction.
F160W contours 0.5 magnitudes apart are overlayed over this image.
North is up and east is to the left.
See Table 1 for more information.
\label{fig:fig1} }
\end{figure*}

\begin{figure*}
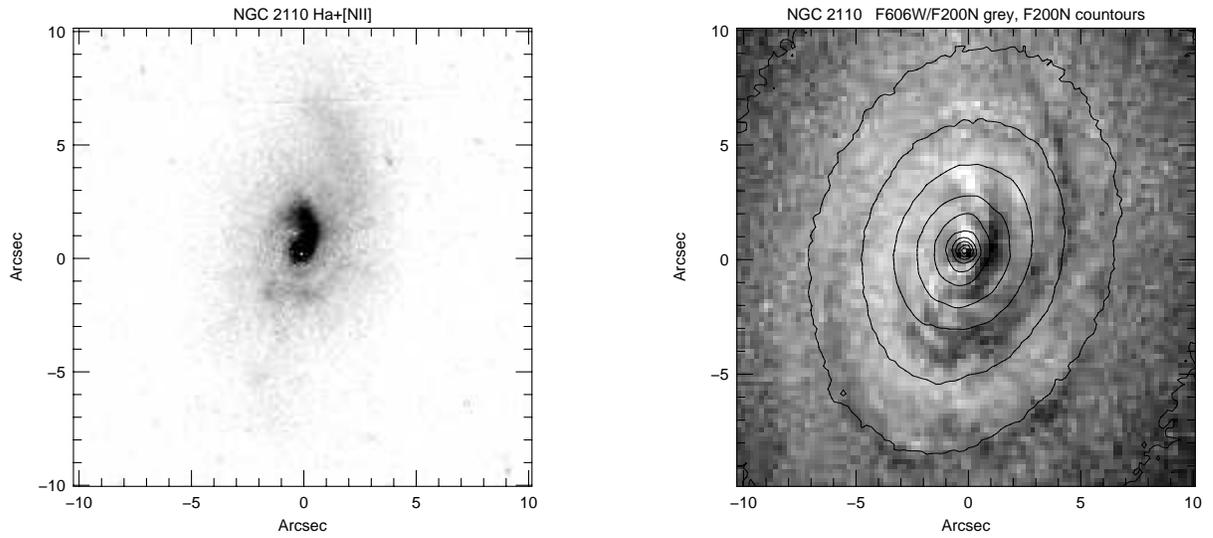

\caption[junk]{
a) The NGC 2110 H$\alpha$+[NII] emission map.
b) The NGC 2110 color map is shown as the greyscale image.
F200N contours 0.5 magnitudes apart are also shown. 
See Fig.~1 and Table 1 for more information.
\label{fig:fig2}}
\end{figure*}

\begin{figure*}
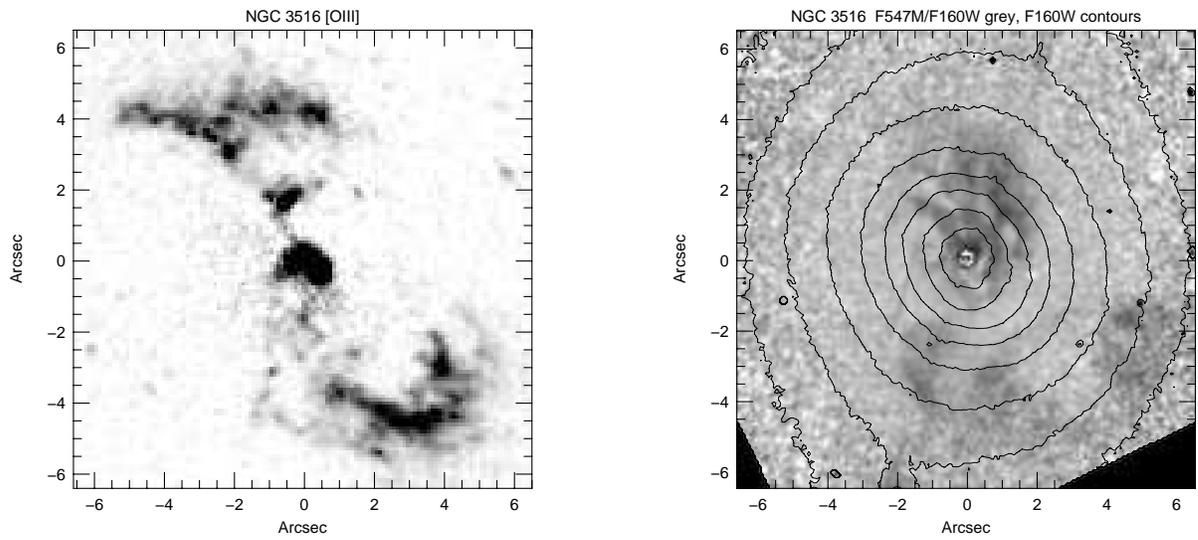

\caption[junk]{
a) The NGC 3516 [OIII] emission map.
b) The NGC 3516 color map is shown as the greyscale image.
F160W contours 0.5 magnitudes apart are also shown.
See Fig.~1 and Table 1 for more information.
\label{fig:fig3} }
\end{figure*}

\begin{figure*}
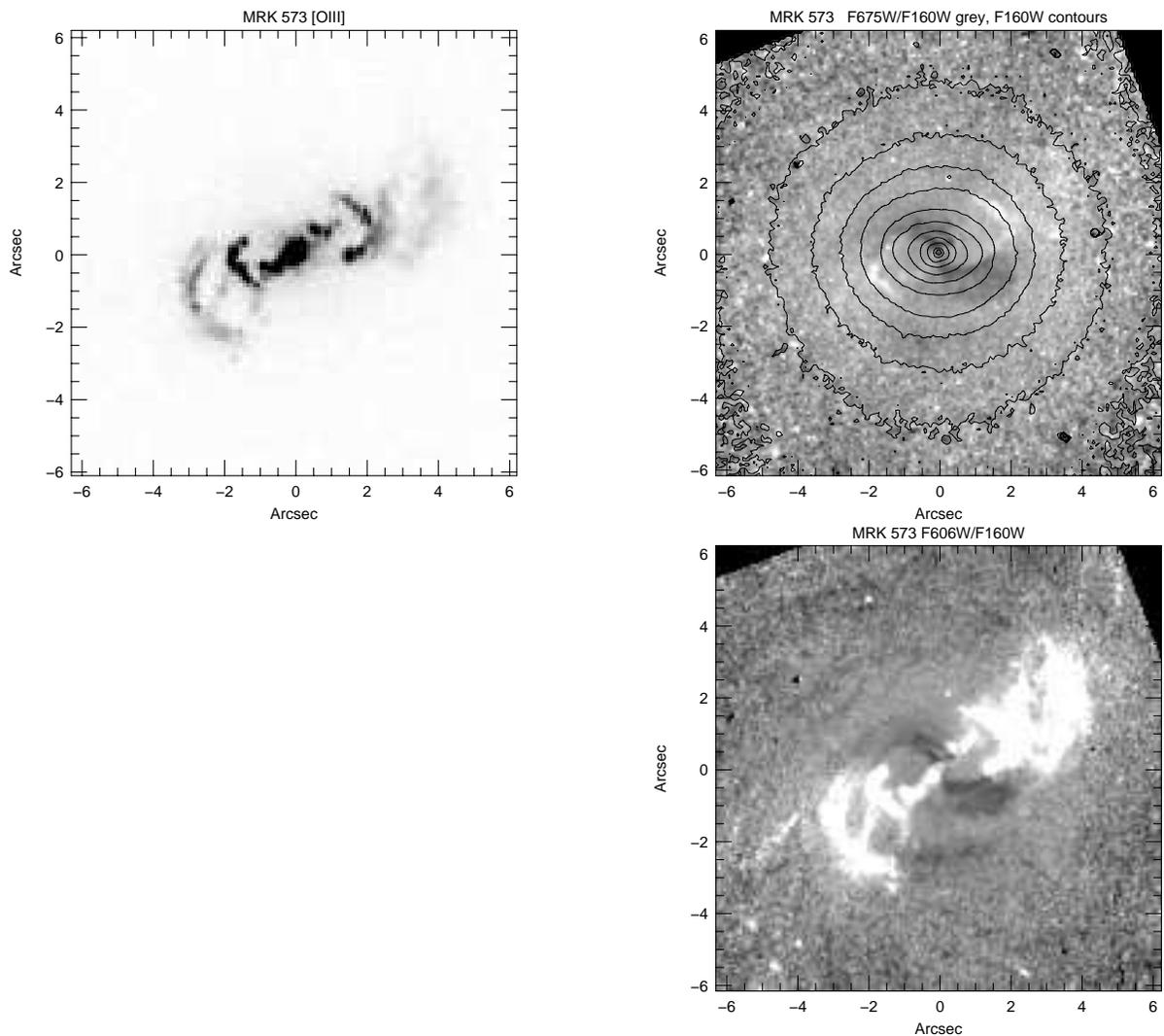

\caption[junk]{
a) The Markarian 573 [OIII] emission map.
b) The Markarian 573 color map is (shown as the greyscale image)
was constructed from the ratio of
F675W and F160W  images.  
Also shown are F160W contours 0.5 magnitudes apart.
c) The color map shown here was constructed from 
from the ratio of F606W and F160W images.
The F606W image is deeper than the F675W image but is contaminated
with line emission.
See Fig.~1 and Table 1 for more information.
\label{fig:fig4} }
\end{figure*}

%\documentstyle[aj_pt4]{article}
%\begin{document}
%\pagestyle{empty}
%\tablenum{1}

% table 1 tab1

\begin{deluxetable}{lcccccccccccc}
\footnotesize
\tablecaption{Images}
%\tablewidth
\tablehead{
\colhead{Galaxy}                           &
\colhead{Scale $1''$}                      &
\colhead{Line Filter}                      &    
\colhead{Continuum Filter}                 &    
\colhead{Visible Filter}                   &    
\colhead{Near-IR Filter}                   &          
\colhead{Max Contour}                      \cr
\multicolumn{1}{c}{(1)} &
\colhead{(2)}                              &
\colhead{(3)}                              &
\colhead{(4)}                              &
\colhead{(5)}                              &
\colhead{(6)}                              &
\colhead{(7)}                               \cr
\colhead{}                                  &
\colhead{(pc)}                                 &                              
\colhead{}                                  &                             
\colhead{}                                  &                             
\colhead{}                                  &                             
\colhead{}                                  &                             
\colhead{(mag/''$^{2}$)}         
} 
\startdata
 NGC 5643 & 77  & F502N  & F547M &  F814W         & F160W & 10.5  \cr
 NGC 2110 & 148 & F664N  & F718M &  F606W         & F200N & 11.0  \cr
 NGC 3516 & 171 & FR533N & F547M &  F547M         & F160W & 13.0  \cr
 MRK 573  & 335 & FR533N &       &  F606W, F675W  & F160W & 12.0  \cr
\enddata
\tablenotetext{}{
Columns-- (1) Galaxy;
(2) Scale in pc for $1''$ assuming $H_0 = 75$km s$^{-1}$ Mpc$^{-1}$;
(3) The [OIII] or H$\alpha$+[NII] emission map was constructed
using this filter containing the emission line; 
(4) The line emission map was constructed using this continuum filter.
Line and continuum images were all observed with WFPC2 except in the case
of NGC~2110 which was observed with the planetary camera;
(5) The WFPC2 image we used to construct color maps displayed
in Figs.~1-4 was observed in this filter.  Central wavelengths are 
F547M ($0.547\mu$m), F606W ($0.60\mu$m) and F675W ($0.675\mu$m);
(6) The NICMOS filter in which the near-infrared
image was observed that we used to construct color maps
and display contours shown in Figs.~1-4,
Central wavelengths are F160W ($1.60\mu$m) and F200N ($2.00\mu$m);
All near-infrared images were taken on NICMOS/Camera~2 except
for that of NGC~2110 which was taken with NICMOS/Camera~3;
(7) Brightest F160W or F200N contour shown in Figs.~1-4.
Calibration reference points were
from M.~Rieke private communication.  The magnitudes in the F160W
filter are very close to Johson H mags.
% The zero point used was 17.94 mag (F200N)
%The zero point used was 21.48 mag (M.~Rieke private communication, F160W).
}
\end{deluxetable}


\begin{thebibliography}{}

%A Near-Infrared Imaging Study of Seyfert Galaxies with Extended 
%Emission-Line Regions
\bibitem[Alonso-Herrero et al.~1998]{almudena}
Alonso-Herrero, A., Simpson, C., Ward, M.~J., \& Wilson, A.~S.\
1998, ApJ, 495, 196

\bibitem[Antonucci 1993]{antonucci}
Antonucci, R.~R.~J.\ 1993, ARA\&A, 31, 473

% dynamics of shocks
\bibitem[Athanassoula 1992]{ath92}
Athanassoula, E.\ 1992, MNRAS,  259, 345

%\bibitem[Baldwin, Wilson, \& Whittle 1987]{baldwin}
%Baldwin, J.~A., Wilson, A.~S., \& Whittle, M.\ 1987,  ApJ, 319, 84

%Faint Object Camera imaging and spectroscopy of NGC 4151
\bibitem[Boksenberg et al.~1995]{boksenberg}  
Boksenberg, A., et al.~1995, ApJ, 440, 151

\bibitem[Dopita \& Sutherland 1995]{dopita}
Dopita, M.~A., \& Sutherland, R.~S.\ 1995, ApJ, 455, 468

\bibitem[Evans et al.~1991]{evans}
Evans, I.~N., Ford, H.~C., Kinney, A.~L., Antonucci, R.~J., 
Armus, L., \& Caganoff, S.\ 1991, ApJ, 369, L27

%HST and VLA observations  of seyfert 2 galaxies 
%the relation between radio ejecta and the NLR
\bibitem[Falcke, Wilson \& Simpson 1998]{falcke3}
Falcke, H., Wilson A.~S., Simpson, C.\ 1998, ApJ, 502, 199

%HST WFPC2 imaging of the seyfert galaxy NGC~3516
\bibitem[Ferruit, Wilson \& Mulchaey 1998]{ferruit98_3}
Ferruit, P., Wilson, A.~S., \& Mulchaey, J.~S.\ 1998, ApJ, 509, 646

%The EELR of the seyfert galaxy Mrk 573
\bibitem[Ferruit et al.~1999]{ferruit99}
Ferruit, P., Wilson, A.~S., Falcke, H., Simpson, C., 
Pecontal, E., \& Durret, F.~1999, MNRAS, in press 

\bibitem[Goad \& Gallagher 1987]{goad}
Goad, J.~W., \& Gallagher, J.~S., III 1987, AJ, 94, 640

%Hydrodynamical simulations of the barred spiral galaxy NGC~1365. 
%Dynamical interpretation of observations.
\bibitem[Lindblad, Lindblad, \&  Athanassoula 1996]{lindblad}
Lindblad, P.~A.~B., Lindblad, P.~O., \& Athanassoula, E.\ 1996, A\&A, 313, 65

\bibitem[Maiolino et al.~1999]{circinus}
Maiolino, R., Alonso-Herrero, A., Rieke, M.~J., Rieke, G.~H.,  \&
Quillen, A.~C.\ 1999, in preparation  

\bibitem[McLeod 1997]{mcl}
McLeod, B.~1997, proceedings of the 1997 HST Calibration Workshop,
eds.~S.~Casertano, R.~Jedrzejewski, T.~Keyes, and M.~Stevens,
published by the Space Telescope Science Institute, Baltimore, MD, p.~281

\bibitem[Morris et al.~1985]{morris}
Morris, S., Ward, M., Whittle, M., Wilson, A.~S., \& Taylor, K.\ 
1985, MNRAS, 216, 193

%The Fueling of Nuclear Activity. 
%I. A Near-Infrared Imaging Survey of Seyfert and Normal Galaxies
\bibitem[Mulchaey, Regan \& Kundu 1997]{regan_3}
Mulchaey, J.~S., Regan, M.~W., \& Kundu, A.\ 1997, ApJS, 110, 299

%
\bibitem[Mulchaey et al.~1992]{mulchaey92}
Mulchaey, J.~S., Tsvetanov, Z., Wilson, A.~S., \& Perez-Fournon, I.\
1992, ApJ, 394, 91 

%An Alignment between Optical Continuum and Emission-Line Structures 
%in the Circumnuclear Regions of Seyfert Galaxies
\bibitem[Mulchaey \& Wilson 1995]{mulchaey95}
Mulchaey, J.~S., \& Wilson, A.~S.\ 1995, ApJ, 455, 17

%HST imaging of sy2 NGC~2110
\bibitem[Mulchaey et al.~1994]{mulchaey94}
Mulchaey, J.~S., Wilson, A.~S., Bower, G.~A., Heckman, T.~M.,
Krolik, J.~H, \& Miley, G.~K.~1994, ApJ, 433, 625

%An Emission-Line Imaging Survey of Early-Type Seyfert Galaxies. 
%II. Implications for Unified Schemes 
%%%%has alot of simulations
\bibitem[Mulchaey, Wilson \& Tsvetanov 1996]{mulchaey96}
Mulchaey, J.~S., Wilson, A.~S., \& Tsvetanov, Z.\
1996, ApJ, 467, 197

%Radio Structures of Seyfert Galaxies. VIII. A Distance and 
%Magnitude Limited Sample of Early-Type Galaxies
\bibitem[Nagar et al.~1999]{nagar} 
Nagar, N.~M., Wilson, A.~S., Mulchaey, J.~S., \& Gallimore, J.~F.\
1999, ApJS, 120, in press, astro-ph/9901214 

\bibitem[Pedlar et al.~1993]{pedlar}
Pedlar, A., et al.\ 1993, MNRAS, 263, 471

\bibitem[Pogge 1988]{pogge}
Pogge, R.~W.\ 1988, ApJ, 332, 702

\bibitem[Pogge 1989]{pogge89}
Pogge, R.~W.\ 1989, ApJ, 345, 730

%Extended near-ultraviolet continuum emission and the nature
%of the polarized broad-line Seyfert 2 galaxies
\bibitem[Pogge \& DeRobertis 1993]{pogge93}
Pogge, R.~W., \& DeRobertis, M.~M.\ 1993, ApJ, 404, 563 

%The Fueling of Active Nuclei: A NICMOS Snapshot Survey of Seyfert 
%and Normal Galaxies
\bibitem[Regan \& Mulchaey]{regan_mul}
Regan, M.~W.,  \& Mulchaey, J.~S.\ 1998, BAAS, 193, 0605

%Spiral structure and star formation. 
%II - Stellar lifetimes and cloud kinematics
\bibitem[Hausman \& Roberts 1984]{hau84}
Hausman, M.~A., \&  Roberts, W.~W., Jr.\  1984, ApJ, 282, 106

%A one sided ionization cone in the seyfert 2 galaxy ngc 5643
\bibitem[Simpson et al.~1997]{simpson}
Simpson, C., Wilson, A.~S., Bower, G. Heckman, T.~M., Krolik, J.~H.,
\& Miley, G.~K.~1997, ApJ, 474, 121

\bibitem[Ulrich \& P\'equignot 1980]{ulrich}
Ulrich, M.~H., \& P\'equignot, D.\ 1980, ApJ, 238, 45

\bibitem[Unger et al.~1987]{unger}
Unger, S.~W., Pedlar, A., Axon, D.~J., Whittle, M., 
Meurs, E.~J.~A., \& Ward, M.~J.\ 1987, MNRAS, 228, 671

\bibitem[Tubbs 1980]{tubbs}
Tubbs, A.~D.\ 1980, ApJ, 241, 969.

\bibitem[Wilson \& Tsvetanov 1994]{wilson+t}
Wilson, A.~S., Tsvetanov, Z.\ 1994, AJ, 107, 1227

%On the origin of the Z-shaped narrow-line region in the Seyfert galaxy NGC 3516
\bibitem[Veilleux, Tully, \& Bland-Hawthorn 1993]{veilleux}
Veilleux, S., Tully, R.~B., Bland-Hawthorn, J.\ 1993, AJ, 105, 1318

\bibitem[Falcke et al.~1998]{falcke}
\bibitem[Ferruit et al.~1998]{ferruit98}
\bibitem[Mulchaey et al.~1997]{regan}

\end{thebibliography}
\end{document}